# From DDMs to DNNs: Using process data and models of decision-making to improve human-AI interactions

Mrugsen Nagsen Gopnarayan[1], *Jaan Aru[2], and Sebastian Gluth*[1]

[1]*Department of Psychology; University of Hamburg*

[2]*Institute of Computer Science; University of Tartu*

Abstract

Over the past decades, cognitive neuroscientists and behavioral economists have recognized the value of describing the process of decision making in detail and modeling the emergence of decisions over time. For example, the time it takes to decide can reveal more about an agent's true hidden preferences than only the decision itself. Similarly, data that track the ongoing decision process such as eye movements or neural recordings contain critical information that can be exploited, even if no decision is made. Here, we argue that artificial intelligence (AI) research would benefit from a stronger focus on insights about how decisions emerge over time and from incorporating related process data to improve AI predictions in general and human AI interactions in particular. First, we discuss to what extent current approaches in multi-agent AI do or do not incorporate process data and models of decision making. Next, we introduce a highly established computational framework that assumes decisions to emerge from the noisy accumulation of evidence, and we present related empirical work in psychology, neuroscience, and economics. Finally, we provide specific examples of how a more principled inclusion of the evidence-accumulation framework into the training and use of AI can help to improve human-AI interactions in the future.

Correspondence concerning this article should be addressed to Mrugsen Nagsen Gopnarayan, E-mail: mrugsen.gopnarayan@uni-hamburg.de or mrugsenng@gmail.com

**Introduction**

From screening mammography for breast cancer tumors with expert-level accuracy (McKinney et al., 2020) to providing more accurate and localized weather forecasts (Sobash et al., 2020), the AI revolution is positively impacting our lives. However, in many of these cases where AI systems perform well, the AI system does not engage in social interaction with a human. Such examples are the solution to the protein folding problem (Jumper et al., 2021), handwriting recognition (Lecun et al., 1998), and the discovery of novel drugs (Zhavoronkov et al., 2019). In these use-cases, humans still set goals by giving labels and setting loss or reward functions, and artificial systems are used as a tool to do an asocial task. Yet, there are many contexts such as geriatric care, psychotherapy, or personal assistance, where social interactions are crucial for achieving favorable outcomes, and an effective human-AI interaction becomes a prerequisite to address the issue. In such instances, AI systems are expected to simulate human roles, assisting or coordinating with the user. A competent artificial agent should, therefore, be capable of understanding human behavior, accounting for the likes, beliefs, intentions, desires, and emotions of the person they interact with. This unique capacity is termed Theory of Mind (ToM). In this review, our goal is to explore existing methods that have been used to try to achieve ToM in AI and to propose a novel approach inspired by decision neuroscience. By focusing on this aspect, we aim to bridge the gap between AI's current capabilities in simpler, nonsocial settings and the demands of socially interactive environments.

In the pursuit of equipping AI with ToM capabilities, lessons can be learned from the functioning of the human brain. This strategy has historically fueled numerous AI breakthroughs. To name just two, the inception of Artificial Neural Networks (ANN) was inspired by McCulloch and Pitts, 1943's model, which mimics biological neurons. Similarly, Convolutional Neural Networks (CNNs) are inspired by the visual processing mechanisms of the human brain, notably how neurons are structured in the visual cortex, tailoring them for image recognition tasks (Lecun et al., 1998). In this paper, we argue that to enhance AI's capability to develop ToM, inspiration can be drawn from how our brain implements ToM in the context of social decision-making. Crucially, some current theories indicate ToM to be implemented in the brain through simulating the process of decision making of others (Gallese, 1998; Gordon, 1986). Hence, we argue that understanding general decisionmaking processes and using process-tracing methods to capture those mechanisms might be the key to ToM in AI.

In the following, we begin by examining the current state of multi-agent AI systems and their applications in complex scenarios, such as multi-player games and autonomous vehicle traffic management. Despite their sophistication, we identify the limitations of these systems in achieving true ToM capability. We then shift to discuss decision-making mechanisms in our brain, and argue why process tracing models are better at capturing the nuances of human decision-making. Following this, we present an overview of experimental techniques used to observe decision-making processes. In the subsequent section, we discuss how process models and data from process tracing methods can be utilized to train Deep Neural Networks (DNNs) for simulating human decision-making. We then highlight specific areas where this approach could be transformative and also where this approach may not work. This is followed by a discussion on trust, ethical implications, and potential challenges in implementing this approach.

**Progress and pitfalls in interactive AI**

When it comes to emulating human social and cooperative behaviors, the classic single-agent machine learning approaches are limited. Conventional single-agent reinforcement learning (RL) approaches such as Q-Learning or policy gradient, for instance, perform poorly as the environments become dynamic and agents change their behavior with training, leading to a moving goal post (Schwartz, 2014). However, inspired by social and behavioral sciences (Duffy et al., 1998) and powered by deep learning, successful multi-agent AI approaches have emerged. These approaches, often termed Multi-Agent Deep Reinforcement Learning (MADRL), involve the use of interconnected DNNs as agents within a system, which are trained using RL principles. Each of these agents interacts with both the environment and other agents in the system, to optimize a set of objectives or rewards. Capable of learning to represent complex functions, predicting future states, or estimating the value of different actions, this approach revolutionized multi-agent AI (Gronauer & Diepold, 2022). These agents can master complex multi-agent video games with superhuman capacity using just simple visual information (Tampuu et al., 2017), and can be adapted to manage the traffic of autonomous vehicles (Zhao et al., 2021).

This level of advancement in MADRL permits the modeling of complex social and cooperative behaviors, akin to those observed in humans and animals (Lowe et al., 2017). Despite the success of AI models such as ToMnet, a DNN that claims to model the agents it encounters solely from observations of their behavior (Rabinowitz et al., 2018), it is debatable whether these networks have truly attained ToM capabilities. For instance, instead of learning the difference between internal states and true states for other agents, an agent like ToMnet might exploit the combination of positions and distances between elements to navigate. This suggests that rather than developing genuine ToM, DNNs may have learned shortcuts, solving a simpler problem than achieving ToM (Aru et al., 2023). Indeed, in humans, ToM is not merely task-based. For example, children do not repetitively learn to solve the Sally-Anne task to gain ToM. Therefore, instead of training deep learning agents on specific tasks that might require ToM, a better approach would be to train them in more complex environments (Aru et al., 2023).

One solution to the problem would be to learn from the experts in ToM, i.e., humans, to acquire this ability. Humans can help agents to learn the policy or reward function, using imitation learning and inverse RL, respectively. For instance, in the realm of autonomous vehicles, where inverse RL is employed, the vehicle learns from observing human drivers (Plebe et al., 2024). But, instead of learning the difference between internal states and true states for other agents, an agent like ToMnet might exploit the combination of positions and distances between elements to navigate. Nevertheless, learning from human preferences (Christiano et al., 2017) or explanation-based learning could bias the agent towards using features with more generalization power and identifying the world's causal structure. For example, the AI model that powers ChatGPT has been significantly enhanced by human feedback loops during its training phase (Ouyang et al., 2022).This iterative feedback process helped shape the model's responses over time, making it more aligned with human values and preferences. ChatGPT correctly solved % of text-based false belief tasks it was tested against (Kosinski, 2023), but see Ullman (2023). This approach can specifically be used to imitate human decision-making processes. AI can gain insights into human preferences, biases, and reasoning strategies, by observing and learning from human decision-making patterns.

In real life, people often need to make decisions for others (like buying a gift for Secret Santa) or with others (such as deciding on a location for a team retreat). We can expand on the idea of learning from humans to train AI, to the domain of social decision-making. Importantly, humans take each other's mental states, preferences, beliefs, intentions, and emotions into account when making decisions in social settings, i.e., they use ToM. One approach to ToM in humans is called 'simulation theory', which suggests that humans simulate others to understand them (Gallese, 1998; Gordon, 1986). Thus, a person A could simulate another person B by replaying (or preplaying) the decision that B has made (or is about to make), using A's own decision-making system. Notably, the discovery of mirror neurons that are active both when we perform an action and when we witness someone else performing the same action has been seen as supporting the simulation account of TOM. (Gordon, 1986; Rizzolatti & Craighero, 2004).

This process tracing is key to our simulation of others. Our core proposal is that implementation of ToM in AI would benefit from learning from process models of decision making, which encapsulate a series of cognitive steps that occur during decision-making, reflecting the dynamic interplay of perception, evaluation, and action. In the subsequent sections, we will briefly explain these models and describe different tools that can be used to track the emergence of decisions over time before addressing the question of how this knowledge can foster AI development.

## Modeling the emergence of decisions using process models

Understanding how choices come about requires process models of decision-making that describe how decisions emerge over time (from sensory input to motor output). In this regard, human decision-making has been modeled as the accumulation of information over time until a decision threshold is met. The drift-diffusion model (DDM) is a widely used mathematical model that represents the accumulation of evidence over time at a specific rate (the drift rate), with some added noise (diffusion)(Ratcliff, 1978; Ratcliff et al., 2016). Although being the dominant model in the field, the DDM is only one representative of the much larger class of evidence accumulation models (EAMs) (Busemeyer et al., 2019; Smith & Ratcliff, 2004). Importantly, parameters of EAMs reflect latent psychological processes. In case of the DDM, for example, a pre-existing bias for either option is modeled by the starting point parameter, the decision boundary represents the cautiousness of the decision maker, while the drift rate models the rate of evidence accumulation. Non-decision time is also considered and assumed to result from sensory-motor processes that are not related to the decision per se. EAMs offers a very general framework for decision-making that can be extended to perceptual tasks (Summerfield & Tsetsos, 2012).

Strikingly, a series of recent studies have shown that laypeople seem to intuitively understand a fundamental prediction of EAMs. Specifically, these studies indicate that the time it takes to make a decision usually reflects the decision's difficulty and, consequently, the difference in preference for the options; fast responses are associated with low difficulty and a high difference in preference (Figure 1). In other words, humans take not only decisions but also decision speed into account when inferring others' hidden preferences and beliefs (Bavard et al., 2023; Konovalov & Krajbich, 2017). The process by which individuals infer others' preferences can be modeled by inverting the DDM and adopting a Bayesian approach (Gates et al., 2021). However, the neural mechanisms of this ability remain elusive.

This complex interplay of theory and simulation in human ToM can also be understood as a form of process tracing imitation. Humans try to simulate scenarios from others' perspectives. This concept provides the motivation for our suggestion that DNNs should be trained to simulate human decisions. These networks could partially gain the human ability to understand and predict others' decisions, thereby enhancing AI's capabilities in social contexts.

## Tracking physiological and neural processes of decisions

Many advanced AI techniques like neural networks have an immense capacity for learning and pattern recognition, but they also require enormous amounts of data to train. Thankfully, neuroscientists have been using a wide array of techniques to effectively learn and infer the process of decision-making. These methods provide a window into the workings of the human brain, offering clues on how decisions emerge and evolve. Hence, using the data obtained from these techniques could be beneficial in training AI systems capable of ToM.

A variety of physiological and neural tools have been used to track the emergence of decisions. Ideally, these measures can be compared with process models of decisionmaking such as the DDM. Although fMRI allows identifying brain regions that are relevant to decision-making, the much higher temporal resolution of Electroencephalography (EEG) and Magnetoencephalography (MEG) make them more promising candidates to track the dynamic emergence of decisions, including sub-processes such as perception, evaluation, and response preparation (Proudfit, 2014). Contrary to prior belief that these processes happen sequentially (Posner, 1986; Sternberg, 1969), it is now well established that these sub-processes overlap in time and space (i.e., brain regions) to a substantial degree (Gluth et al., 2012; Hare et al., 2011). There is a tight coupling between accumulated evidence, as modeled by EAMs and the activity in the pre-supplementary motor area (pre-SMA). The Readiness Potential (RP) is an event-related EEG potential that emerges in this region while we are preparing an action (Kornhuber & der Deecke, 1965; Shibasaki & Hallett, 2006). The RP can be understood as a "tendency to respond", and an experimentally induced sense of urgency leads to an increase in the signal (Gluth et al., 2013). The lateralized RP is another neural marker that can be used to investigate the dynamics of action selection. In a task where the choice has to be made using two different hands, LRPs can be calculated as the difference in potentials for left responses vs. right responses over the right vs. left primary motor cortices. Notably, the onset of the LRP could be an indicator of different non-decision times in decisions with different cognitive demands (memory-based vs. regular decisions) (Kraemer & Gluth, 2023). In perceptual decision making, the centro-parietal positivity (CPP) has been proposed as a marker of evidence accumulation (O'Connell et al., 2012), though this claim has recently been challenged (Frömer et al., 2022).

Besides neural mechanisms, there are also powerful tools to track peripheral and physiological signals that are intertwined with decision-making processes. Among these, eye-tracking is particularly relevant, since eye movements serve as a window into a decisionmaker's attentional processes. Eye movements seem to both influence and reflect preference (Shimojo et al., 2003) and have been used to inform EAMs. The attentional Drift-Diffusion Model (aDDM) (Krajbich et al., 2010) assumes that fixated options impact the accumulation process more than non-fixated options. Importantly, many studies have shown that taking eye-movements into account via the aDDM (or similar models) improves predicting decisions substantially (Gluth et al., 2018, 2020; Krajbich et al., 2010). In addition to eye movements, pupil responses are another critical physiological data source as pupil dilation appears to be a reliable measure of arousal (Joshi et al., 2016), and reveals how decisions evolve (de Gee et al., 2014). Relatedly, choices made contrary to (and thus overcoming) default responses lead to an increased pupil dilation. The starting point in DDM, which represents such response biases, is predictive of pupil dilation (Sheng et al., 2020). Mouse tracking can also reflect the decision process. When individuals encounter conflicting options or experience decision difficulty, their mouse movements can exhibit curvilinear or wavering trajectories, reflecting internal conflict or deliberation (Spivey & Dale, 2006). Heart rate, skin conductance, and cortisol levels are other physiological methods for understanding the decision-making process. However, these measures are more indirect and have high latency, thus they are not widely used in process-tracing studies.

Although seemingly distinct, social decisions significantly overlap with individual decisions with respect to involved brain regions. Relatedly, it has been argued and shown that the basic framework of conceptualizing value-based choicess as emerging from evidence accumulation also applies to social decision-making (Gluth & Fontanesi, 2016; Hu et al., 2023; Hunt & Hayden, 2017). EAMs can explain choices and response times in social decision making tasks as well. Moreover, simultaneous EEG-fMRI recordings have shown that the process of evidence accumulation in the medial prefrontal cortex (mPFC), which includes the pre-SMA, is comparable across social and non-social decisions (Arabadzhiyska et al., 2022). However, compared to individual decision-making, social decision-making is arguably more complex, as it involves understanding and taking others' preferences or biases into account, and learning from others. As a consequence, additional brain regions appear to be critical to social decision making, many of them being said to belong to the ToM network. Central to this process is the temporoparietal junction (TPJ), a key component of the Theory of Mind network. The TPJ plays a critical role in differentiating one's perspective from others', especially in bargaining situations, and in processing social information and integrating fairness norms during decision-making (Hu et al., 2023; Saxe & Kanwisher, 2003). Obviously, humans cannot directly access the neurophysiological information of others when interacting with them. Instead, they need to infer others' thoughts and intentions by observing their behavior, body language, facial expressions, and gaze patterns. Eye movements are especially important as they serve as a window to a decision maker's thought process, revealing their preferences and influencing the outcomes in social interactions. Eye-tracking studies have shown that displaying the gaze allocation of one participant to another in a coordination game improves understanding of each other's preferred choices and facilitates strategic decision-making for maximizing rewards (Hausfeld et al., 2020).

**Improving human-AI interactions with process models of decision-making**

As discussed above, significant advances have been made to comprehend the computational, physiological, and neural principles of decision-making processes in both individual and social contexts (Konovalov & Ruff, 2021). EAMs have been instrumental in this journey, coupled with various tools such as fMRI, EEG/MEG, and eye-tracking. Recent studies, which focused on how people interpret the decisions of others, have started to shed light on an intuitive human understanding of decision processes that align with EAM principles (Gates et al., 2021). In the following, we build on this work to outline several ideas of how to implement these insights into multi-agent AI systems with DNNs. These ideas are thought to help materialize the theoretical understanding of human decision-making processes into tangible AI applications and could help mimicking ToM capabilities in AI. We will structure our ideas with specific steps involved in the implementation of process learning in DNN:

1. **Data**

Until now, mostly choice data has been used to train neural networks (Kuperwajs et al., 2023; Peterson et al., 2021). But as discussed in the previous sections, there is a much richer set of data available that contains information about the decision-making process which has not been used thus far. DNNs could also be fed with response times, EEG data, gaze patterns, pupil dilation, or other process tracing datasets. This will help DNNs to not only mimic the choices made but also the process that resulted in these choices. These datasets may be pre-processed, as is common in most analysis approaches of neural and physiological data in neuroscience and psychology. But preprocessing the data might actually lead to a loss of information, and hence should be avoided whenever possible (Khosla et al., 2022). Nevertheless, datasets should be properly formatted, for instance, by converting stimulus into a format that can be processed by a neural network. Additionally, data simulated using process tracing models could also be used to train the neural network. For example, simulated data using DDMs has been used to train an RNN to match the decision time and the final choice in a probabilistic learning task (Zhang et al., 2020).

## 2. Model Architecture

Popular architectures for modeling decisions are recurrent neural networks (RNNs) which excel at handling sequential data, making them suitable for modeling decision-making processes that involve a series of steps. Within the family of RNNs, Long Short-Term Memory (LSTM) networks and Gated Recurrent Units (GRUs) are particularly effective. They can remember information over longer periods and are less prone to issues like vanishing gradients, which can be a problem with standard RNNs (Shewalkar et al., 2019). Another way of tailoring the architecture is to create a multi-domain network, a network comprising different types of layers. For example, a variational layer can be used to capture the inherent variability and uncertainty in human decision-making. This layer often employs techniques from variational inference (Oleksiienko et al., 2023). It can represent the probability distributions of various latent variables that influence decision-making, such as risk preference or uncertainty in outcomes. An attention layer, incorporating attention mechanisms, like those used in transformers, can also be beneficial (Vaswani et al., 2017). These mechanisms help the model focus on the most relevant parts of the input data, which is similar to how humans focus more on important details. CNNs can also be particularly useful in the pre-processing stage, especially when the input data includes spatial or visual elements. CNNs are adept at extracting hierarchical features from images or spatial data. They use convolutional layers to apply filters that detect patterns, edges, and other relevant features.

Tailoring the architecture of the DNN to handle the specifics of process-tracing is crucial. One way to achieve this is by mathematically constraining nodes of the neural networks, so that they are functionally equivalent to decision making models. This has been done where different decision making models were compared against one another by representing them as neural networks (See Peterson et al., 2021).

## 3. Training

Training: The training of the network with process tracing data is the key step. This might involve a mix of both supervised and unsupervised learning. Supervised Learning could be helpful for mapping specific stimuli and physiological data (e.g. sensory stimulus, gaze data, EEG data) to decision-making outcomes (e.g., final decisions, response times), or DDN can be asked to predict the physiological data, ensuring they would also learn about the process of decision making (Figure 2A). Unsupervised Learning, particularly reinforcement learning, could be applied to tasks that involve learning from environmental interactions to maximize cumulative rewards (like in probabilistic learning tasks).

4. Validation

The aim of the validation step is to assess how well the neural network replicates human thought processes in decision-making. This involves evaluating the model's performance against real human behavior in similar scenarios. One aspect of this will be to test the model's outputs by comparing them with decisions made by humans in identical or very similar decision-making tasks. This could involve using a separate validation dataset that the model has not been trained on. Another aspect should be to test the AI's performance in novel and distinct choice environments and interactive settings, to test how well the system can generalize to these novel instances. Importantly, such generalization tests may also allow us to identify whether a multi-agent AI system has obtained proper ToM abilities or simply learned shortcuts within a narrow range of choice problems.

### Real-World Implications of Process Tracing approach in AI

Given the intricate nature of human-AI interactions, the ability of AI systems to understand, predict, and adapt to human decision-making processes is not just an advantage but a prerequisite for their effective integration into our lives. In the last section we discussed how process tracing data and models can be used to train an AI - an approach that could enable AI to mirror and understand the complexities of human decision-making. These models, rooted in Theory of Mind (ToM), will allow AI systems to anticipate human preferences, beliefs, and intentions, with more speed and precision, thereby fostering more natural and effective interactions. In this section, we explore practical scenarios in detail, where we expect the incorporation of process tracing models into AI to significantly enhance its social intelligence and to improve human-AI interaction.

**E-Commerce Mouse Tracing**

An e-commerce website could utilize mouse tracing to enhance user experience and recommendation accuracy. The website could analyze how users navigate through the site, observing patterns such as the speed of mouse movement, areas where the cursor lingers, and items receiving more clicks. Based on these analyses, the site could infer user interests and indecisions. For instance, if a user frequently hovers over a particular category without making a purchase, the website could deduce a high interest but some level of hesitation, possibly leading to targeted discounts or suggestions of similar but different products to aid in the decision-making process.

**House Help Robot**

In the future, a robot designed to assist with daily chores, such as cooking or ordering groceries, could use process tracing models to learn from the past decisions of household members. The robot could also interpret the decision-making process, where quick choices might suggest a strong preference for certain dishes, and more time spent pondering could indicate a desire for variety(Figure 2B). This learning could enable the robot to make dinner decisions that closely align with the family's preferences.

**Supermarket Vending Machines**

Supermarkets could be equipped with smart vending machines that recommend products to customers based on process tracing. These machines could analyze customers' eye movements as they look at different products, interpreting patterns such as the length of time a customer's gaze lingers on certain items or how their eyes move between different product types. This data could help the machines understand customer preferences and indecision, allowing them to make personalized recommendations. For example, if a customer frequently pauses at healthy snacks but usually chooses chocolates, the machine could suggest a sugar free chocolate bar that combines elements of both.

Besides these encouraging possibilities for enhanced human-AI interactions, there are also scenarios where this approach may be less fruitful. Take, for instance, the field of therapeutics. Human emotions and mental health issues are incredibly complex and varied. Process tracing models, while sophisticated, might still fall short of fully grasping and appropriately responding to the wide range of human emotional experiences, especially in nuanced therapeutic contexts. Similarly, in the realm of autonomous driving, process models still do not seem to be well poised to capture complicated human behaviors involved in driving, and hence may have limited use. However, analyzing the hesitation of drivers before making turns or pedestrians crossing streets, could significantly augment autonomous driving systems.

## Challenges and future directions

Integrating AI systems into our daily lives comes with significant challenges, notably the lack of transparency and interpretability of AI systems. A core problem in this regard is "trustworthy AI", including questions such as whether and under what circumstances people feel comfortable interacting with AI systems, or whether people are willing to transfer responsibilities to machines. Human users often find it difficult to comprehend how an AI system arrives at its decisions, and this lack of transparency leads to reduced trust and acceptance. The major reason contributing to algorithm aversion is the "black box" nature of these systems (Mahmud et al., 2022). When users do not clearly understand how an algorithm works and how it generates decisions, they tend to exhibit aversion towards these systems, as evidenced by studies like (Dzindolet et al., 2002). People often need justification for algorithmic decisions, and the absence of understandable explanations reduces their trust in these systems (Lu et al., 2017). Hence, people naturally prefer human decisionmakers as they can ask them for an explanation behind the decision. The issue of trust is further exacerbated when there is cognitive dissonance due to nonconforming decisions from algorithms (Kitayama et al., 2013).

To elevate this problem, providing clear and understandable explanations of decisions is crucial. When users receive decisions accompanied by explanations of how the algorithm works, they perceive it as more trustworthy (Goodwin et al., 2013). Furthermore, the work of (Lankton et al., 2015) draws attention to the intricate relationship between technology, humanness, and trust, highlighting that AI systems imbued with human-like qualities can foster a higher degree of trust among users. Our proposal to incorporate process tracing models into AI systems can harness this to address the challenge of trust. By mirroring human decision-making processes, these models could offer a more human-like and interpretable framework for AI systems, potentially improving user trust and acceptance. This could be combined with other explainable AI techniques, such as generating understandable rules or highlighting influential factors. This has shown promise in improving human understanding of AI decisions (Kraus et al., 2020). Effective communication strategies, such as using personalized conversation, illustrations, and persuasive language, can also enhance the perception of trust and acceptance (Yun et al., 2021).

Another challenge in using this framework is the ethical consideration in collecting human data for process training approaches. This approach heavily relies on extensive multi-modal data on human behavior. However, the collection and use of such data raise critical ethical questions regarding privacy, consent, and the potential misuse of personal information. Ensuring that data collection adheres to strict ethical standards and respects individual privacy rights is imperative. This involves transparent data collection practices, securing informed consent, and implementing robust data protection measures.

Although this approach of process tracing is quite promising, it necessitates experimental validation to assess its efficacy thoroughly. Initial laboratory studies can be conducted where AI systems are trained using this framework and subsequently evaluated against other approaches. This phase of experimentation is critical to establish the reliability of the process training approach. Another vital expansion would be the incorporation of models that can capture emotional states. This could be achieved by integrating data from facial expressions, body language, and even physiological signals. By training neural networks to interpret these subtle cues, AI systems can gain a deeper understanding of the emotional context behind human decisions. Applying process tracing models to more complex scenarios is another critical direction. For instance, in the context of autonomous driving, AI systems can benefit from understanding the decision-making processes of human drivers, such as reaction times, attention allocation, and risk assessment strategies. Finally, continuous improvement and adaptation of process tracing models is key. As our understanding of human behavior and decision-making evolves, so too should the models we use to train AI systems.

In conclusion, an integration of decision-making research into multi-agent AI development holds promising potential for enhancing human-AI interaction. Leveraging process models of human decision-making within AI systems can create agents that more accurately reflect and predict human behavior, thus fostering better mutual understanding. These advances can enable the creation of AI systems that are not only more effective, but also more transparent and trustworthy. As we continue this exploration, our focus should be on ensuring AI systems that are not just intelligent, but also relatable and acceptable to their human users, thereby paving the way for a more symbiotic relationship between humans and AI.

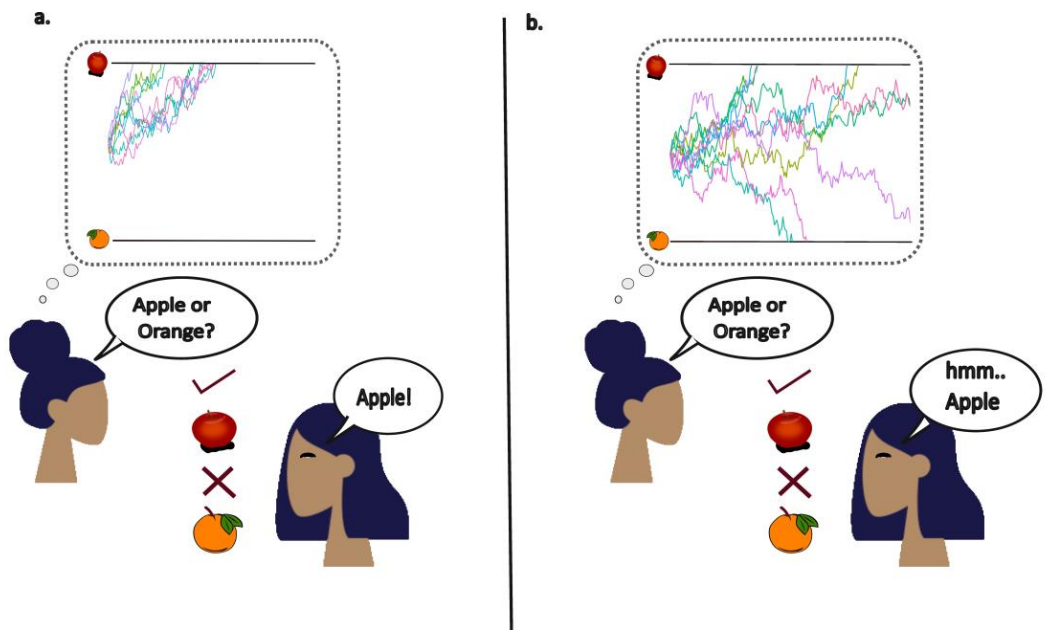

**Figure 1**

*Simulation Theory in Inferring Preferences. The figure demonstrates how humans simulate others' decisions to infer preferences. In (a), a rapid choice for an apple over an orange implies a strong preference for apples. Conversely, in (b), a more deliberative process with multiple simulated scenarios, not all leading to the choice of an apple, suggests weak preference. The decision trajectories symbolize the cognitive simulation process done by an observer, highlighting how decision latency and pattern can be indicative of preference strength.*

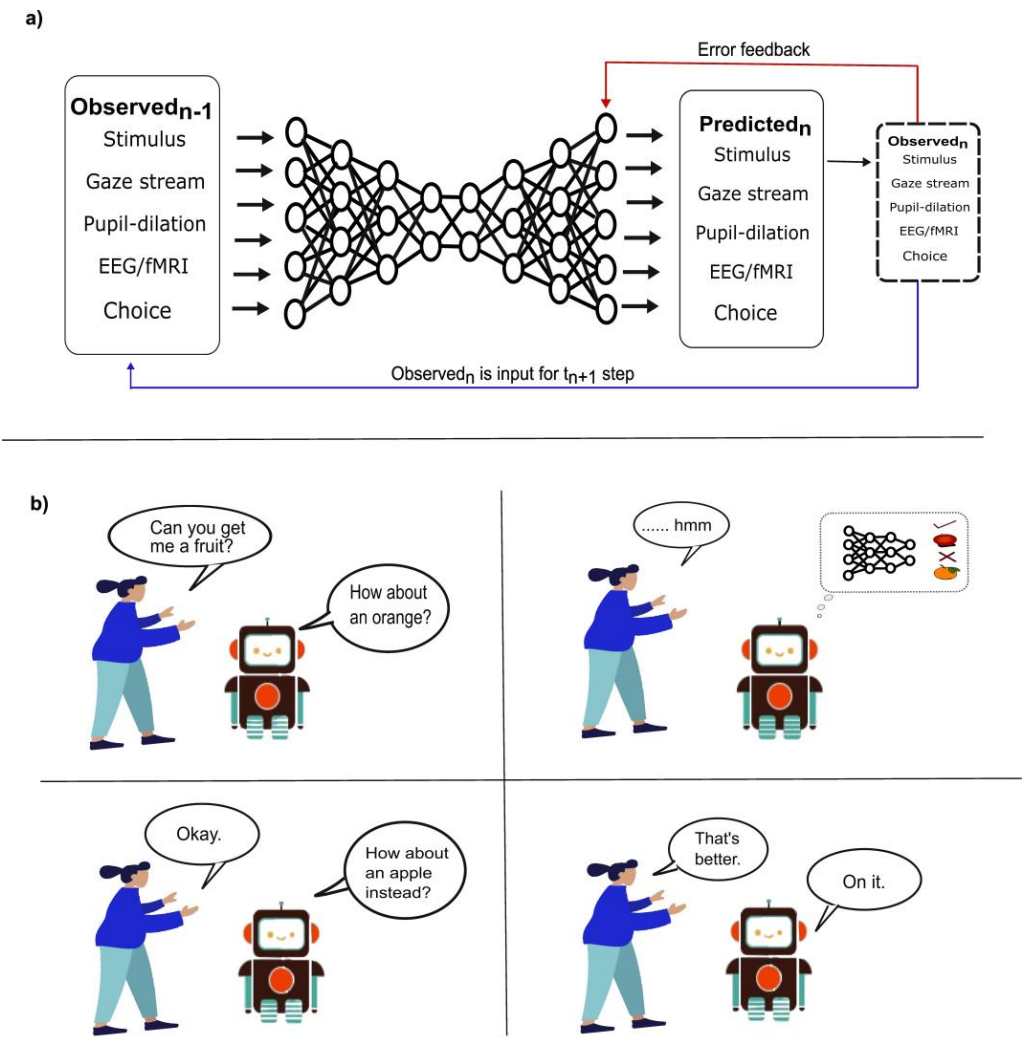

**Figure 2**

*Deep Neural Network (DNN) Employing a Process Tracing Training Mechanism. a)The figure illustrates a method of training deep neural networks. On the left, Observed(n-1) provides the neural network with historical data, including stimulus, gaze stream, pupil dilation, EEG/fMRI, and choice, which are used as inputs for training. The network generates predictions of the same features, which are then compared against the Observed(n) data, represented on the right. The comparison yields an error feedback, denoted by the red arrow, which is fed back into the network to adjust and improve the model's predictive capabilities.*

*b) A person requests a fruit, and the AI robot uses a neural network having a process tracing mechanism to infer from the person's hesitation that its first suggestion (orange) was not good enough (as indicated by the thought bubble). Using the same network, the AI then identifies a*

*better alternative (apple). This would show AI's ability to interpret and respond to human preferences.*

Gluth, S., Spektor, M. S., & Rieskamp, J. (2018). Value-based attentional capture affects multi-alternative decision making. *eLife*, *7*. https://doi.org/10.7554/elife.39659

Goodwin, P., Sinan Gönül, M., & Önkal, D. (2013). Antecedents and effects of trust in forecasting advice. *International Journal of Forecasting*, *29*(2), 354–366. https://doi.org/10.1016/j.ijforecast.2012.08.001

Gordon, R. M. (1986). Folk psychology as simulation. *Mind and Language*, *1*(2), 158–71. https://doi.org/10.1111/j.1468-0017.1986.tb00324.x

Gronauer, S., & Diepold, K. (2022). Multi-agent deep reinforcement learning: A survey. *Artificial Intelligence Review*, 1–49.

Hare, T. A., Malmaud, J., & Rangel, A. (2011). Focusing attention on the health aspects of foods changes value signals in vmPFC and improves dietary choice. *Journal of Neuroscience*, *31*(30), 11077–11087. https://doi.org/10.1523/jneurosci.6383-10.2011

Hausfeld, J., von Hesler, K., & Goldlücke, S. (2020). Strategic gaze: An interactive eyetracking study. *Experimental Economics*, *24*(1), 177–205. https://doi.org/10.1007/s10683-020-09655-x

Hu, J., Konovalov, A., & Ruff, C. C. (2023). A unified neural account of contextual and individual differences in altruism. *eLife*, *12*. https://doi.org/10.7554/elife.80667

Hunt, L. T., & Hayden, B. Y. (2017). A distributed, hierarchical and recurrent framework for reward-based choice. *Nature Reviews Neuroscience*, *18*(3), 172–182. https://doi.org/10.1038/nrn.2017.7

Joshi, S., Li, Y., Kalwani, R. M., & Gold, J. I. (2016). Relationships between pupil diameter and neuronal activity in the locus coeruleus, colliculi, and cingulate cortex. *Neuron*, *89*(1), 221–234. https://doi.org/10.1016/j.neuron.2015.11.028

Jumper, J., Evans, R., Pritzel, A., Green, T., Figurnov, M., Ronneberger, O., Tunyasuvunakool, K., Bates, R., Žídek, A., Potapenko, A., Bridgland, A., Meyer, C., Kohl, S. A. A., Ballard, A. J., Cowie, A., Romera-Paredes, B., Nikolov, S., Jain, R., Adler, J., ... Hassabis, D. (2021). Highly accurate protein structure prediction with AlphaFold. *Nature*, *596*(7873), 583–589. https://doi.org/10.1038/s41586-021-03819-2

Khosla, A., Khandnor, P., & Chand, T. (2022). Automated diagnosis of depression from eeg signals using traditional and deep learning approaches: A comparative analysis. *Biocybernetics and Biomedical Engineering*, *42*(1), 108–142. https://doi.org/10.1016/j.bbe.2021.12.005

Kitayama, S., Chua, H. F., Tompson, S., & Han, S. (2013). Neural mechanisms of dissonance: An fmri investigation of choice justification. *NeuroImage*, *69*, 206–212. https://doi.org/10.1016/j.neuroimage.2012.11.034

Konovalov, A., & Krajbich, I. (2017). On the strategic use of response times. *SSRN Electronic Journal*. https://doi.org/10.2139/ssrn.3023640

Konovalov, A., & Ruff, C. C. (2021). Enhancing models of social and strategic decision making with process tracing and neural data. *WIREs Cognitive Science*, *13*(1). https://doi.org/10.1002/wcs.1559

Kornhuber, H. H., & der Deecke, L. (1965). Hirnpotential nderungen bei willk rbewegungen und passiven bewegungen des menschen: Bereitschaftspotential und reafferente potentiale. *Pfl gers Archiv f r die Gesamte Physiologie des Menschen und der Tiere*, *284*(1), 1–17. https://doi.org/10.1007/bf00412364

Kosinski, M. (2023). Theory of mind may have spontaneously emerged in large language models. *arXiv preprint arXiv:2302.02083*.

FROM DDM TO DNN    15Kraemer, P. M., & Gluth, S. (2023). Episodic memory retrieval affects the onset and dynamics of evidence accumulation during value-based decisions. *Journal of Cognitive Neuroscience*, *35*(4), 692–714. https://doi.org/10.1162/jocn_a_01968

Krajbich, I., Armel, C., & Rangel, A. (2010). Visual fixations and the computation and comparison of value in simple choice. *Nature Neuroscience*, *13*(10), 1292–1298. https://doi.org/10.1038/nn.2635

Kraus, S., Azaria, A., Fiosina, J., Greve, M., Hazon, N., Kolbe, L., Lembcke, T.-B., Muller, J. P., Schleibaum, S., & Vollrath, M. (2020). Ai for explaining decisions in multiagent environments. *Proceedings of the AAAI conference on artificial intelligence*, *34*(09), 13534–13538.

Kuperwajs, I., Schütt, H. H., & Ma, W. J. (2023). Using deep neural networks as a guide for modeling human planning. *Scientific Reports*, *13*(1). https://doi.org/10.1038/s41598-023-46850-1

Lankton, N., McKnight, D. H., & Tripp, J. (2015). Technology, humanness, and trust: Rethinking trust in technology. *Journal of the Association for Information Systems*, *16*(10), 880–918. https://doi.org/10.17705/1jais.00411

Lecun, Y., Bottou, L., Bengio, Y., & Haffner, P. (1998). Gradient-based learning applied to document recognition. *Proceedings of the IEEE*, *86*(11), 2278–2324. https://doi.org/10.1109/5.726791

Lowe, R., Wu, Y. I., Tamar, A., Harb, J., Pieter Abbeel, O., & Mordatch, I. (2017). Multiagent actor-critic for mixed cooperative-competitive environments. *Advances in neural information processing systems*, *30*.

Lu, J., Liang, Y., & Duan, H. (2017). Justifying decisions: Making choices for others enhances preferences for impoverished options. *Social Psychology*, *48*(2), 92–103. https://doi.org/10.1027/1864-9335/a000302

Mahmud, H., Islam, A. N., Ahmed, S. I., & Smolander, K. (2022). What influences algorithmic decision-making? a systematic literature review on algorithm aversion. *Technological Forecasting and Social Change*, *175*, 121390. https://doi.org/10.1016/j.techfore.2021.121390

McCulloch, W. S., & Pitts, W. (1943). A logical calculus of the ideas immanent in nervous activity. *The Bulletin of Mathematical Biophysics*, *5*(4), 115–133. https://doi.org/10.1007/bf02478259

McKinney, S. M., Sieniek, M., Godbole, V., Godwin, J., Antropova, N., Ashrafian, H., Back, T., Chesus, M., Corrado, G. S., Darzi, A., Etemadi, M., Garcia-Vicente, F., Gilbert, F. J., Halling-Brown, M., Hassabis, D., Jansen, S., Karthikesalingam, A., Kelly, C. J., King, D., ... Shetty, S. (2020). International evaluation of an AI system for breast cancer screening. *Nature*, *577*(7788), 89–94. https://doi.org/10.1038/s41586-019-1799-6

O'Connell, R. G., Dockree, P. M., & Kelly, S. P. (2012). A supramodal accumulation-to-bound signal that determines perceptual decisions in humans. *Nature Neuroscience*, *15*(12), 1729–1735. https://doi.org/10.1038/nn.3248

Oleksiienko, I., Tran, D. T., & Iosifidis, A. (2023). Variational neural networks [International Neural Network Society Workshop on Deep Learning Innovations and Applications (INNS DLIA 2023)]. *Procedia Computer Science*, *222*, 104–113. https://doi.org/ https://doi.org/10.1016/j.procs.2023.08.148

FROM DDM TO DNN 17Spivey, M. J., & Dale, R. (2006). Continuous dynamics in real-time cognition. *Current Directions in Psychological Science*, *15*(5), 207–211. https://doi.org/10.1111/j.14678721.2006.00437.x

Sternberg, S. (1969). The discovery of processing stages: Extensions of donders' method. *Acta Psychologica*, *30*, 276–315. https://doi.org/10.1016/0001-6918(69)90055-9

Summerfield, C., & Tsetsos, K. (2012). Building bridges between perceptual and economic decision-making: Neural and computational mechanisms. *Frontiers in Neuroscience*, *6*. https://doi.org/10.3389/fnins.2012.00070

Tampuu, A., Matiisen, T., Kodelja, D., Kuzovkin, I., Korjus, K., Aru, J., Aru, J., & Vicente, R. (2017). Multiagent cooperation and competition with deep reinforcement learning (C.-Y. Xia, Ed.). *PLOS ONE*, *12*(4), e0172395. https://doi.org/10.1371/journal.pone.0172395

Ullman, T. (2023). Large language models fail on trivial alterations to theory-of-mind tasks. *arXiv preprint arXiv:2302.08399*.

Vaswani, A., Shazeer, N., Parmar, N., Uszkoreit, J., Jones, L., Gomez, A. N., Kaiser, Ł., & Polosukhin, I. (2017). Attention is all you need. In I. Guyon, U. V. Luxburg, S. Bengio, H. Wallach, R. Fergus, S. Vishwanathan, & R. Garnett (Eds.), *Advances in neural information processing systems* (Vol. 30). Curran Associates, Inc. https://proceedings.neurips.cc/paper_files/paper/2017/file/3f5ee243547dee91fbd053c1c4a845aa-Paper.pdf

Yun, J. H., Lee, E.-J., & Kim, D. H. (2021). Behavioral and neural evidence on consumer responses to human doctors and medical artificial intelligence. *Psychology amp; Marketing*, *38*(4), 610–625. https://doi.org/10.1002/mar.21445

Zhang, Z., Cheng, H., & Yang, T. (2020). A recurrent neural network framework for flexible and adaptive decision making based on sequence learning (A. Soltani, Ed.). *PLOS Computational Biology*, *16*(11), e1008342. https://doi.org/10.1371/journal.pcbi.1008342

Zhao, C., Li, L., Pei, X., Li, Z., Wang, F.-Y., & Wu, X. (2021). A comparative study of state-of-the-art driving strategies for autonomous vehicles. *Accident Analysis & Prevention*, *150*, 105937. https://doi.org/10.1016/j.aap.2020.105937

Zhavoronkov, A., Ivanenkov, Y. A., Aliper, A., Veselov, M. S., Aladinskiy, V. A., Aladinskaya, A. V., Terentiev, V. A., Polykovskiy, D. A., Kuznetsov, M. D., Asadulaev, A., Volkov, Y., Zholus, A., Shayakhmetov, R. R., Zhebrak, A., Minaeva, L. I., Zagribelnyy, B. A., Lee, L. H., Soll, R., Madge, D., ... Aspuru-Guzik, A. (2019). Deep learning enables rapid identification of potent DDR1 kinase inhibitors. *Nature Biotechnology*, *37*(9), 1038–1040. https://doi.org/10.1038/s41587-019-0224-x